\documentclass{optica-article}

\journal{oe}

\articletype{Research Article}

\begin{document}
\title{Enantiodiscrimination of chiral molecules via quantum correlation function}

\author{Fen Zou\authormark{1}, Yu-Yuan Chen\authormark{1,2}, Bo Liu\authormark{1}, and Yong Li\authormark{3,1,4,$\ast$}}

\address{\authormark{1}Beijing Computational Science Research Center, Beijing 100193, China}
\address{\authormark{2}School of Integrated Circuits, Tsinghua University, Beijing 100084, China}
\address{\authormark{3}Center for Theoretical Physics and School of Science, Hainan University, Haikou 570228, China}
\address{\authormark{4}Synergetic Innovation Center for Quantum Effects and Applications, Hunan Normal University, Changsha 410081, China}

\email{\authormark{$\ast$}yongli@hainanu.edu.cn} 

\begin{abstract}
We propose a method to realize enantiodiscrimination of chiral molecules based on quantum correlation function in a driven cavity-molecule system, where the chiral molecule is coupled with a quantized cavity field and two classical light fields to form a cyclic three-level model. According to the inherent properties of electric-dipole transition moments of chiral molecules, there is a $\pi$-phase difference in the overall phase of the cyclic three-level model for the left- and right-handed chiral molecules. Thus, the correlation function depends on this overall phase and is chirality-dependent. The analytical and numerical results indicate that the left- and right-handed chiral molecules can be discriminated by detecting quantum correlation function. Our work opens up a promising route to discriminate molecular chirality, which is an extremely important task in pharmacology and biochemistry.
\end{abstract}

\section{Introduction}
Chiral molecules cannot be superimposed with its mirror image by translation and rotation~\cite{Woolley1976Quantum}. Due to the fact that the left- and right-handed chiral molecules (called enantiomers) are mirror images of each other, they have similar physical properties. However, there are significant differences in the physiological effects for the enantiomers of biologically active compounds~\cite{Franks1991Stereospecific,Bodenhofer1997Chiral}. Thus, enantiodiscrimination~\cite{Jia2011Probing,Hirota2012Triple,Yachmenev2016Detecting,Lehmann2018Influence,Ye2019Determination,Chen2020Enantio,
Xu2020Enantiomeric,Kang2020Effective,Chen2021Enantio,Ye2021Entanglement,Chen2021arXivEnantio,Cai2021arXivEnantiodetection}, enantiospecific state transfer~\cite{Kral2001Cyclic,Li2008Dynamic,Jia2010Distinguishing,Leibscher2019Principles,Ye2019Effective,
Vitanov2019Highly,Wu2019Robust,Wu2020Two,Wu2020Discrimination,Torosov2020Efficient,Torosov2020Chiral,Zhang2020Evading}, spatial enantioseparation~\cite{Li2007Generalized,Li2010Theory,Jacob2012Effect,Eilam2013Spatial,Bradshaw2015Laser,Liu2021Spatial,Suzuki2019Stern,
Milner2019Controlled}, and enantioconversion~\cite{Shapiro2000Coherently,Brumer2001Principles,Gerbasi2001Theory,Kral2003Two,Frishman2004Optical,Ye2020Fast,Ye2021An,
Ye2021Enantio} of chiral molecules have become essential issues. So far, the traditional spectroscopic methods, such as optical rotation~\cite{Kondru1998Atomic}, circular dichroism~\cite{Beaulieu2000Photoexcitation}, vibrational circular dichroism~\cite{Stephens1985Theory}, and Raman optical activity~\cite{Bielski2005Absolute}, have been proposed to detect the molecular chirality. Nevertheless, it should be pointed out that these spectroscopic methods are based on the interference between the electric- and magnetic-dipole (or electric-quadrupole) transitions. Since the magnetic-dipole and electric-quadrupole transition moments are usually weak compared with the electric-dipole transition moment, the chiral signal obtained by the spectroscopic methods is relatively weak. This indicates that enantiodiscrimination of chiral molecules remains a challenging task.

In recent years, some theoretical schemes have been proposed to implement the enantiodiscrimination~\cite{Jia2011Probing,Hirota2012Triple,Yachmenev2016Detecting,Lehmann2018Influence,Ye2019Determination,Chen2020Enantio,
Xu2020Enantiomeric,Kang2020Effective,Chen2021Enantio,Ye2021Entanglement,Chen2021arXivEnantio,Cai2021arXivEnantiodetection}, enantiopurification (including enantiospecific state transfer~\cite{Kral2001Cyclic,Li2008Dynamic,Jia2010Distinguishing,Leibscher2019Principles,Ye2019Effective,
Vitanov2019Highly,Wu2019Robust,Wu2020Two,Wu2020Discrimination,Torosov2020Efficient,Torosov2020Chiral,Zhang2020Evading} and enantioconversion~\cite{Shapiro2000Coherently,Brumer2001Principles,Gerbasi2001Theory,Kral2003Two,Frishman2004Optical,Ye2020Fast,Ye2021An,
Ye2021Enantio}), and spatial enantioseparation~\cite{Li2007Generalized,Li2010Theory,Jacob2012Effect,Eilam2013Spatial,Bradshaw2015Laser,
Liu2021Spatial} of chiral molecules through the optical means. In particular, the experiments on enantiodiscrimination~\cite{Patterson2013Enantiomer,Patterson2013Sensitive,
Patterson2014New,Shubert2014Identifying,Shubert2015Rotational,Shubert2016Chiral,Lobsiger2015Molecular} and enantiospecific state transfer~\cite{Eibenberger2017Enantiomer,Perez2017Coherent} have been achieved based on cyclic three-level models of chiral molecules in the microwave regime~\cite{Kral2001Cyclic,Kral2003Two,Liu2015Optical,Ye2018Real}. In such three-level models, the product of the three corresponding coupling strengths of the electric-dipole transition moments changes sign with enantiomers. Therefore, there is a $\pi$-phase difference in the overall phase of the cyclic three-level model for the left- and right-handed chiral molecules~\cite{Kral2001Cyclic,Kral2003Two}. By exploiting this property of the cyclic three-level model, enantiodiscrimination of the chiral molecules can be realized. However, most of previous theoretical and experimental studies on enantiodiscrimination of the chiral molecules focused on the researches of the classical physical quantities, e.g. the optical absorption spectra~\cite{Jia2011Probing}, the population difference~\cite{Kral2001Cyclic,Li2008Dynamic,Jia2010Distinguishing,Leibscher2019Principles,Ye2019Effective,
Vitanov2019Highly,Wu2019Robust,Wu2020Two,Wu2020Discrimination,Torosov2020Efficient,Torosov2020Chiral,Zhang2020Evading,Eibenberger2017Enantiomer,Perez2017Coherent}, the deflection angle of the light~\cite{Chen2020Enantio}, the intensity of the output field~\cite{Chen2021Enantio}, the transmission rate~\cite{Chen2021arXivEnantio}, etc.

In this paper, we propose a method for the discrimination of the left- and right-handed chiral molecules based on quantum correlation function (i.e., equal-time second-order correlation function) of the cavity field. The second-order correlation function is a quantum physical quantity and can reveal the quantum properties of the field. By analysing the second-order correlation function in cyclic three-level models of atom~\cite{Li2019Enhanced,wu2021Phase}, the photon (or magnon) blockade effect has been investigated. Here, based on the similar cyclic three-level models of chiral molecules, we study enantiodiscrimination of chiral molecules by detecting the equal-time second-order correlation function. In the weak-driving case, we derive the analytical expression of the equal-time second-order correlation function by using the probability amplitude method~\cite{Bamba2011Origin,Xu2014Strong,Xu2016Phonon,Bin2018Two,Zou2020Enhancement}, and find that the correlation function is chirality-dependent. According to the theoretical analysis, we further demonstrate that the left- and right-handed chiral molecules can be distinguished by detecting the correlation function. Therefore, our method provides a feasible way to discriminate molecular chirality.

The rest of this paper is organized as follows. In Sec.~\ref{modelsec}, we introduce the physical model of the cavity-molecule system and present the system Hamiltonian. In Sec.~\ref{corrfunsec}, we derive the analytical expression of the equal-time second-order correlation function of the cavity field in the weak-driving case. In Sec.~\ref{detchirsec}, we investigate the dependence of the equal-time second-order correlation function for the left- and right-handed chiral molecules on the parameters (e.g. the detunings and the driving strengths). Finally, A summary is given in Sec.~\ref{conclusion}.
\begin{figure}
\center
\includegraphics[width=0.65 \textwidth]{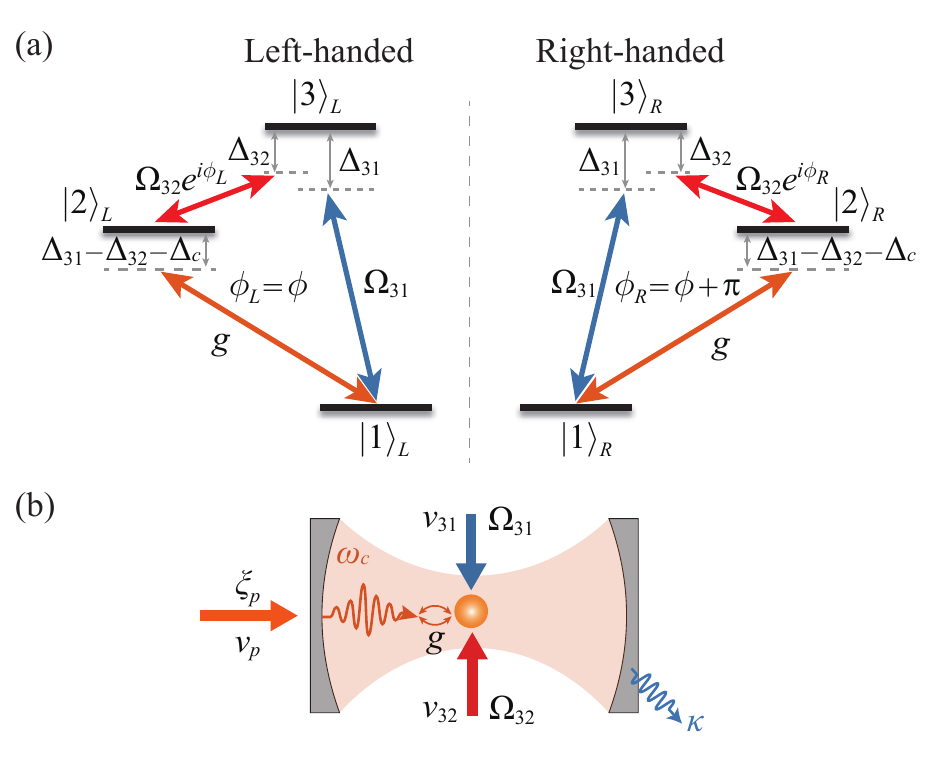}
\caption{(a) Schematic diagram of the cyclic three-level left- and right-handed chiral molecules, where the chiral molecule is coupled with a quantized cavity field and two classical light fields ($\Omega_{31}$ and $\Omega_{32}$). The overall phase $\phi_{Q}$ of the cyclic three-level model for the left- and right-handed chiral molecules is chirality-dependent: $\phi_{R}=\phi_{L}+\pi$. (b) Schematic diagram of the cavity-molecule system consisting of a cavity and a chiral molecule. A weak field $\xi_{p}$ with driving frequency $\nu_{p}$ is applied to the cavity, and two classical light fields $\Omega_{31}$ and $\Omega_{32}$ with driving frequency $\nu_{31}$ and $\nu_{32}$ drive the transitions $\vert1\rangle_{Q}\leftrightarrow\vert3\rangle_{Q}$ and $\vert2\rangle_{Q}\leftrightarrow\vert3\rangle_{Q}$ ($Q=L$ or $R$) of the chiral molecule, respectively.}
\label{Fig1}
\end{figure}

\section{Model\label{modelsec}}
We consider a cavity-molecule system consisting of a cavity and a chiral molecule, as shown in Fig.~\ref{Fig1}. The chiral molecule is coupled with a quantized cavity field and two classical light fields to form the cyclic three-level model, and the cavity is continuously driven by a monochromatic weak field with the driving strength $\xi_{p}$ and driving frequency $\nu_{p}$. For
convenience, we adopt the subscripts ``$L$" and ``$R$" to mark the left- and right-handed chiral molecules, respectively. $\vert j\rangle_{L}$ and $\vert j\rangle_{R}$ ($j=1,2,3$) are, respectively, the $j$th eigen-states of the left- and right-handed chiral molecules with the same eigen-energy $\hbar\omega_{j}$. Under the dipole approximation and rotating-wave approximation, the Hamiltonian of the cavity-molecule system for the left- or right-handed chiral molecules reads ($\hbar=1$)
\begin{align}\label{Hams}
H^{Q} & =\omega_{c}a^{\dagger}a+\sum_{j=1}^{3}\omega_{j}\sigma_{jj}^{Q}
+\left[ga^{\dagger}\sigma_{12}^{Q}+\xi_{p}ae^{i\nu_{p}t}+\Omega_{31}\sigma_{13}^{Q}e^{i\nu_{31}t}+\Omega_{32}e^{i\phi_{Q}}\sigma_{23}^{Q}e^{i\nu_{32}t}
+\text{H.c.}\right],
\end{align}
where the superscript $Q$ ($Q=L$ or $R$) is introduced to represent the molecular chirality. Here $a^{\dagger}$ ($a$) is the creation (annihilation) operator of the cavity field with resonance frequency $\omega_{c}$, and $\sigma_{ij}^{Q}=\vert i\rangle_{Q}\,_{Q}\langle j\vert$ $(i,j=1,2,3)$ are the molecular raising and lowering operators for $i\neq j$ and the molecular population operators for $i=j$. The parameter $g$ denotes the coupling strength between the molecular transition $\vert1\rangle_{Q}\leftrightarrow\vert2\rangle_{Q}$ and the quantized cavity field, $\Omega_{31}$ ($\Omega_{32}$) and $\nu_{31}$ ($\nu_{32}$) are the driving strength and driving frequency of the classical light field acting on the molecular transition $\vert1\rangle_{Q}\leftrightarrow\vert3\rangle_{Q}$ ($\vert2\rangle_{Q}\leftrightarrow\vert3\rangle_{Q}$), and $\phi_{Q}$ is overall phase of the cyclic three-level model. For simplicity but without loss of generality, we have taken these parameters ($g,\xi_{p},\Omega_{31}$, and $\Omega_{32}$) as positive real numbers. The molecular chirality is reflected by choosing the overall phases of the left- and right-handed chiral molecules as~\cite{Kral2001Cyclic,Kral2003Two}
\begin{equation}
\phi_{L}=\phi,\quad \phi_{R}=\phi+\pi.
\end{equation}

In the following discussion, we consider the three-photon resonant condition, i.e., $\nu_{31}=\nu_{32}+\nu_{p}$. In the interaction picture with respect to $H_{0}^{Q}=\omega_{1}\sigma_{11}^{Q}+(\omega_{1}+\nu_{31}-\nu_{32})\sigma_{22}^{Q}+(\omega_{1}+\nu_{31})\sigma_{33}^{Q}+\nu_{p}a^{\dagger}a$, Hamiltonian~(\ref{Hams}) can be written as
\begin{align}\label{HamR}
H_{I}^{Q} & =\Delta_{c}a^{\dagger}a+(\Delta_{31}-\Delta_{32})\sigma_{22}^{Q}+\Delta_{31}\sigma_{33}^{Q}
+\left(ga^{\dagger}\sigma_{12}^{Q}+\xi_{p}a+\Omega_{31}\sigma_{13}^{Q}+\Omega_{32}e^{i\phi_{Q}}\sigma_{23}^{Q}
+\text{H.c.}\right),
\end{align}
where the detunings are defined as $\Delta_{c}=\omega_{c}-\nu_{p}$, $\Delta_{31}=\omega_{3}-\omega_{1}-\nu_{31}$, and $\Delta_{32}=\omega_{3}-\omega_{2}-\nu_{32}$.

In general, the dynamics of the system can be described by quantum master equation~\cite{Kang2020Effective,scully1997Quantum}
\begin{align}\label{MEQ}
\frac{d\rho^{Q}}{dt} & =-i[H_{I}^{Q},\rho^{Q}]+\frac{\kappa}{2}\mathcal{L}_{a}[\rho^{Q}]+\frac{\gamma_{31}}{2}\mathcal{L}_{\sigma_{13}^{Q}}[\rho^{Q}]
+\frac{\gamma_{32}}{2}\mathcal{L}_{\sigma_{23}^{Q}}[\rho^{Q}]+\frac{\gamma_{21}}{2}\mathcal{L}_{\sigma_{12}^{Q}}[\rho^{Q}]\nonumber\\
&\quad+\frac{\gamma_{\phi,31}}{2}\mathcal{L}_{\sigma_{z,31}^{Q}}[\rho^{Q}]+\frac{\gamma_{\phi,32}}{2}\mathcal{L}_{\sigma_{z,32}^{Q}}[\rho^{Q}]+\frac{\gamma_{\phi,21}}{2}\mathcal{L}_{\sigma_{z,21}^{Q}}[\rho^{Q}],
\end{align}
where $\rho^{Q}$ ($Q=L$ or $R$) is the density operator of the cavity-molecule system, $\kappa$ is the decay rate of the cavity field, and $\gamma_{ij}$ ($i>j$ and $i,j=1,2,3$) is the relaxation rate of the chiral molecule from state $\vert i\rangle_{Q}$ to $\vert j\rangle_{Q}$. $\mathcal{L}_{o}[\rho^{Q}]=2o\rho^{Q}o^{\dagger}-o^{\dagger}o\rho^{Q}-\rho^{Q}o^{\dagger}o$ denotes the Lindblad superoperator with $o=a$, $\sigma_{13}^{Q}$, $\sigma_{23}^{Q}$, and $\sigma_{12}^{Q}$. In addition, $\gamma_{\phi,ij}$ is the pure dephasing rate associated with the operator $\sigma_{z,ij}^{Q}=\vert i\rangle_{Q}\,_{Q}\langle i\vert-\vert j\rangle_{Q}\,_{Q}\langle j\vert$. By using the Python package QuTiP~\cite{Johansson2012QuTiP,Johansson2013QuTiP} to numerically solve Eq.~(\ref{MEQ}), we can obtain the density operator $\rho_{\text{ss}}^{Q}$ ($Q=L,R$) of the system in the steady-state case. Here the subscript ``$\text{ss}$" denotes the steady state of the system. Hence, the equal-time second-order correlation function of the cavity field~\cite{scully1997Quantum} can be obtained by $g^{(2)}_{Q}(0)\equiv\langle a^{\dagger2}a^{2}\rangle_{\text{ss}}/\langle a^{\dagger}a\rangle_{\text{ss}}^{2}=\text{Tr}(a^{\dagger2}a^{2}\rho_{\text{ss}}^{Q})/[\text{Tr}(a^{\dagger}a\rho_{\text{ss}}^{Q})]^{2}$. Typically, the quantum correlation function can be measured by employing the Hanbury-Brown-Twiss setup~\cite{Birnbaum2005Photon,Faraon2008Coherent} or quadrature amplitude detectors~\cite{Bozyigit2011Antibunching,Lang2011Observation} in experiments.

\section{Analytical results of the correlation function\label{corrfunsec}}

In order to obtain the analytical results of the correlation function in the cavity field, we consider a weak-driving case  $(\{\Omega_{31},\xi_{p}\}\ll\kappa)$. In this case, the two driving terms $\Omega_{31}(\sigma_{13}^{Q}+\sigma_{31}^{Q})$ and $\xi_{p}(a+a^{\dagger})$ can be considered as perturbative terms. For the Hamiltonian $H_{I}^{\prime Q}=\Delta_{c}a^{\dagger}a+(\Delta_{31}-\Delta_{32})\sigma_{22}^{Q}+\Delta_{31}\sigma_{33}^{Q}
+(ga^{\dagger}\sigma_{12}^{Q}+\Omega_{32}e^{i\phi_{Q}}\sigma_{23}^{Q}+\text{H.c.})$ in the absence of the two driving terms, the total excitation number operator $\hat{N}^{Q}=a^{\dagger}a+\sigma_{22}^{Q}+\sigma_{33}^{Q}$ is a conserved quantity due to the commutative relation $[\hat{N}^{Q},H_{I}^{\prime Q}]=0$. The subspaces corresponding to the total excitation number $N^{Q}=0,1,2,\ldots,n,\ldots$ are spanned over the basis states $\{\vert1,0\rangle_{Q}\}$, $\{\vert1,1\rangle_{Q},\vert2,0\rangle_{Q},\vert3,0\rangle_{Q}\}$, $\{\vert1,2\rangle_{Q},\vert2,1\rangle_{Q},\vert3,1\rangle_{Q}\}$, $\ldots$, $\{\vert1,n\rangle_{Q},\vert2,n-1\rangle_{Q},\vert3,n-1\rangle_{Q}\}$, $\ldots$, where $\vert j,n\rangle_{Q}=\vert j\rangle_{Q}\otimes\vert n\rangle$ defines the state with the chiral molecule in the state of $\vert j\rangle_{Q}$ ($j=1,2,3$) and $n$ ($n=0,1,2,\ldots$) photons in the cavity mode.

To include the influence of the dissipations of the cavity field and the chiral molecule on the quantum statistics, we phenomenologically add the imaginary dissipation terms to Hamiltonian~(\ref{HamR}) as~\cite{Li2019Enhanced,wu2021Phase}
\begin{equation}\label{nonHam}
H_{\text{non}}^{Q}=H_{I}^{Q}-i\frac{\kappa}{2}a^{\dagger}a-i\frac{\gamma_{21}}{2}\sigma_{22}^{Q}-i\left(\frac{\gamma_{31}}{2}
+\frac{\gamma_{32}}{2}\right)\sigma_{33}^{Q}
\end{equation}
with $\kappa$ and $\gamma_{ij}$ ($i,j=1,2,3$) being the decay rates of the cavity field and the chiral molecule, respectively. Here the non-Hermitian Hamiltonian $H_{\text{non}}^{Q}$ is obtained based on the quantum-jump approach~\cite{Plenio1998The} and the pure dephasing of the system is neglected.

In the weak-driving case $(\{\Omega_{31},\xi_{p}\}\ll\kappa)$, we can truncate the Hilbert space of the cavity field up to $n=2$, a general state of the system then can be written as~\cite{Bamba2011Origin,Xu2014Strong,Xu2016Phonon,Bin2018Two,Zou2020Enhancement}
\begin{align}\label{psit}
\vert\psi(t)\rangle_{Q}&=C^{Q}_{1,0}(t)\vert1,0\rangle_{Q}+C^{Q}_{1,1}(t)\vert1,1\rangle_{Q}+C^{Q}_{2,0}(t)\vert2,0\rangle_{Q}+C^{Q}_{3,0}(t)\vert3,0\rangle_{Q} \nonumber \\
&\quad+C^{Q}_{1,2}(t)\vert1,2\rangle_{Q}+C^{Q}_{2,1}(t)\vert2,1\rangle_{Q}+C^{Q}_{3,1}(t)\vert3,1\rangle_{Q},
\end{align}
where $C^{Q}_{j,n}(t)$ ($j=1,2,3$ and $n=0,1,2$) represents the probability amplitude of the corresponding state $\vert j,n\rangle_{Q}$. Based on the Schr\"{o}dinger equation $i\partial_{t}\vert\psi(t)\rangle_{Q}=H_{\text{non}}^{Q}\vert\psi(t)\rangle_{Q}$, we can obtain the equations of motion for these probability amplitudes $C^{Q}_{j,n}(t)$ as
\begin{align}\label{Cjn}
i\dot{C}^{Q}_{1,0} & =\xi_{p}C^{Q}_{1,1}+\Omega_{31}C^{Q}_{3,0},\nonumber \\
i\dot{C}^{Q}_{1,1} & =\delta_{c}C^{Q}_{1,1}+gC^{Q}_{2,0}+\xi_{p}C^{Q}_{1,0}+\sqrt{2}\xi_{p}C^{Q}_{1,2}+\Omega_{31}C^{Q}_{3,1},\nonumber \\
i\dot{C}^{Q}_{2,0} & =\delta_{1}C^{Q}_{2,0}+gC^{Q}_{1,1}+\xi_{p}C^{Q}_{2,1}+\Omega_{32}e^{i\phi_{Q}}C^{Q}_{3,0},\nonumber \\
i\dot{C}^{Q}_{3,0} & =\delta_{2}C^{Q}_{3,0}+\xi_{p}C^{Q}_{3,1}+\Omega_{31}C^{Q}_{1,0}+\Omega_{32}e^{-i\phi_{Q}}C^{Q}_{2,0},\nonumber \\
i\dot{C}^{Q}_{1,2} & =2\delta_{c}C^{Q}_{1,2}+\sqrt{2}gC^{Q}_{2,1}+\sqrt{2}\xi_{p}C^{Q}_{1,1},\nonumber \\
i\dot{C}^{Q}_{2,1} & =(\delta_{1}+\delta_{c})C^{Q}_{2,1}+\sqrt{2}gC^{Q}_{1,2}+\xi_{p}C^{Q}_{2,0}+\Omega_{32}e^{i\phi_{Q}}C^{Q}_{3,1},\nonumber \\
i\dot{C}^{Q}_{3,1} & =(\delta_{2}+\delta_{c})C^{Q}_{3,1}+\xi_{p}C^{Q}_{3,0}+\Omega_{31}C^{Q}_{1,1}+\Omega_{32}e^{-i\phi_{Q}}C^{Q}_{2,1}.
\end{align}
Here we have introduced the parameters $\delta_{c}=\Delta_{c}-i\kappa/2$, $\delta_{1}=\Delta_{31}-\Delta_{32}-i\gamma_{21}/2$, and $\delta_{2}=\Delta_{31}-i(\gamma_{31}+\gamma_{32})/2$.

Under the weak-driving condition ($\{\Omega_{31},\xi_{p}\}\ll\kappa$), there is an approximation scale $C^{Q}_{1,0}\sim1$, $\{C^{Q}_{1,1},C^{Q}_{2,0},C^{Q}_{3,0}\}\sim\xi_{p}/\kappa$, and $\{C^{Q}_{1,2},C^{Q}_{2,1},C^{Q}_{3,1}\}\sim\xi_{p}^{2}/\kappa^{2}$~\cite{Li2019Enhanced,wu2021Phase,Zou2020Enhancement}, namely,
\begin{equation}\label{relaship}
C^{Q}_{1,0}\gg\{C^{Q}_{1,1},C^{Q}_{2,0},C^{Q}_{3,0}\}\gg\{C^{Q}_{1,2},C^{Q}_{2,1},C^{Q}_{3,1}\}.
\end{equation}
In this case, Eq.~(\ref{Cjn}) can be approximately written as
\begin{align}
i\dot{C}^{Q}_{1,0} & \approx0,\nonumber \\
i\dot{C}^{Q}_{1,1} & \approx\delta_{c}C^{Q}_{1,1}+gC^{Q}_{2,0}+\xi_{p}C^{Q}_{1,0},\nonumber \\
i\dot{C}^{Q}_{2,0} & \approx\delta_{1}C^{Q}_{2,0}+gC^{Q}_{1,1}+\Omega_{32}e^{i\phi_{Q}}C^{Q}_{3,0},\nonumber \\
i\dot{C}^{Q}_{3,0} & \approx\delta_{2}C^{Q}_{3,0}+\Omega_{31}C^{Q}_{1,0}+\Omega_{32}e^{-i\phi_{Q}}C^{Q}_{2,0},\nonumber \\
i\dot{C}^{Q}_{1,2} & =2\delta_{c}C^{Q}_{1,2}+\sqrt{2}gC^{Q}_{2,1}+\sqrt{2}\xi_{p}C^{Q}_{1,1},\nonumber \\
i\dot{C}^{Q}_{2,1} & =(\delta_{1}+\delta_{c})C^{Q}_{2,1}+\sqrt{2}gC^{Q}_{1,2}+\xi_{p}C^{Q}_{2,0}+\Omega_{32}e^{i\phi_{Q}}C^{Q}_{3,1},\nonumber \\
i\dot{C}^{Q}_{3,1} & =(\delta_{2}+\delta_{c})C^{Q}_{3,1}+\xi_{p}C^{Q}_{3,0}+\Omega_{31}C^{Q}_{1,1}+\Omega_{32}e^{-i\phi_{Q}}C^{Q}_{2,1},
\end{align}
where we have reasonably discarded the higher-order terms in the equations of motion for the lower-order variables.

Assuming that the probability amplitude of the state $\vert 1,0\rangle_{Q}$ is $C^{Q}_{1,0}=1$, then the steady-state solutions of these probability amplitudes can be obtained as
\begin{align}\label{amplit}
C^{Q}_{1,0} & =1, \nonumber\\
C^{Q}_{1,1} & =(\delta_{1}\delta_{2}\xi_{p}+e^{i\phi_{Q}}g\Omega_{31}\Omega_{32}-\xi_{p}\Omega_{32}^{2})W^{-1}, \nonumber\\
C^{Q}_{2,0} & =-(g\delta_{2}\xi_{p}+e^{i\phi_{Q}}\delta_{c}\Omega_{31}\Omega_{32})W^{-1}, \nonumber\\
C^{Q}_{3,0} & =(\delta_{1}\delta_{c}\Omega_{31}+e^{-i\phi_{Q}}g\xi_{p}\Omega_{32}-g^{2}\Omega_{31})W^{-1},\nonumber\\
C^{Q}_{1,2} & =\{\xi_{p}[C^{Q}_{2,0}g-C^{Q}_{1,1}(\delta_{1}+\delta_{c})](\delta_{2}+\delta_{c})+\xi_{p}C^{Q}_{1,1}\Omega_{32}^{2}\nonumber\\
&\quad-e^{i\phi_{Q}}g\Omega_{32}(C^{Q}_{3,0}\xi_{p}+C^{Q}_{1,1}\Omega_{31})\}(\sqrt{2}V)^{-1}, \nonumber\\
C^{Q}_{2,1} & =[e^{i\phi_{Q}}\delta_{c}\Omega_{32}(C^{Q}_{3,0}\xi_{p}+C^{Q}_{1,1}\Omega_{31})+(\delta_{2}+\delta_{c})(C^{Q}_{1,1}g-C^{Q}_{2,0}\delta_{c})\xi_{p}]V^{-1}, \nonumber\\
C^{Q}_{3,1} & =\{[g^{2}-\delta_{c}(\delta_{1}+\delta_{c})](C^{Q}_{3,0}\xi_{p}+C^{Q}_{1,1}\Omega_{31})-e^{-i\phi_{Q}}\xi_{p}\Omega_{32}(C^{Q}_{1,1}g-C^{Q}_{2,0}\delta_{c})\}V^{-1},
\end{align}
where we have introduced the variables
\begin{align}
W&=g^{2}\delta_{2}-\delta_{1}\delta_{2}\delta_{c}+\delta_{c}\Omega_{32}^{2},\nonumber\\
V&=(\delta_{2}+\delta_{c})[\delta_{c}(\delta_{1}+\delta_{c})-g^{2}]-\delta_{c}\Omega_{32}^{2}.
\end{align}
Note that the probability amplitudes $C^{Q}_{j,n}$ in Eq.~(\ref{amplit}) are not normalized. The normalized probability amplitudes are $\mathcal{C}^{Q}_{j,n}\equiv C^{Q}_{j,n}/\mathcal{N}^{Q}$ with the normalization constant $\mathcal{N}^{Q}=(\vert C^{Q}_{1,0}\vert^{2}+\vert C^{Q}_{1,1}\vert^{2}+\vert C^{Q}_{2,0}\vert^{2}+\vert C^{Q}_{3,0}\vert^{2}+\vert C^{Q}_{1,2}\vert^{2}+\vert C^{Q}_{2,1}\vert^{2}+\vert C^{Q}_{3,1}\vert^{2})^{1/2}$.

Therefore, the equal-time second-order correlation function $g_{Q}^{(2)}(0)$ of the cavity field can be expressed by using Eq.~(\ref{relaship}) as
\begin{equation}\label{corrfun}
g^{(2)}_{Q}(0)=\frac{\sum_{j,n}n(n-1)\vert \mathcal{C}^{Q}_{j,n}\vert^{2}}{(\sum_{j,n}n\vert \mathcal{C}^{Q}_{j,n}\vert^{2})^{2}}\approx\frac{2\vert \mathcal{C}^{Q}_{1,2}\vert^{2}}{\vert \mathcal{C}^{Q}_{1,1}\vert^{4}}.
\end{equation}
Here the coefficients $\vert\mathcal{C}^{Q}_{1,1}\vert^{2}=:P_{\vert 1,1\rangle_{Q}}$ and $\vert\mathcal{C}^{Q}_{1,2}\vert^{2}=:P_{\vert 1,2\rangle_{Q}}$ express the state occupations of $\vert 1,1\rangle_{Q}$ and $\vert 1,2\rangle_{Q}$, respectively. Upon inserting the normalized probability amplitudes $\mathcal{C}^{Q}_{1,1}$ and $\mathcal{C}^{Q}_{1,2}$ into Eq.~(\ref{corrfun}), we can obtain the analytical expression of the correlation function $g^{(2)}_{Q}(0)$, which depends on the overall phase $\phi_{Q}$ of the cyclic three-level model. This means that the left- and right-handed chiral molecules can be discriminated in principle by detecting the equal-time second-order correlation function of the cavity field.
\begin{figure}
\center
\includegraphics[width=0.65 \textwidth]{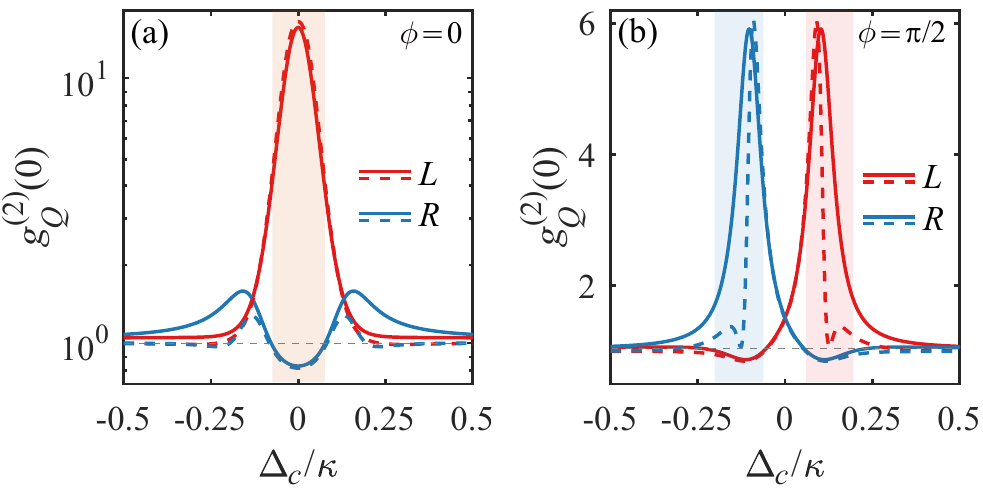}
\caption{The equal-time second-order correlation function $g^{(2)}_{Q}(0)$ ($Q=L,R$) as a function of the detuning $\Delta_{c}$ at (a) $\phi=0$ and (b) $\phi=\pi/2$. The solid and dashed curves correspond to the numerical and analytical results, respectively. Other parameters are chosen as $\Delta_{32}=0$, $\Delta_{31}=\Delta_{c}$, $\kappa/2\pi=1$\,MHz, $\gamma_{21}/2\pi=\gamma_{31}/2\pi=\gamma_{32}/2\pi=0.1$\,MHz, $\gamma_{\phi,21}=\gamma_{\phi,31}=\gamma_{\phi,32}=0$, $g/2\pi=0.1$\,MHz, $\Omega_{32}/2\pi=0.1$\,MHz, and $\xi_{p}/2\pi=\Omega_{31}/2\pi=0.01$\,MHz.}
\label{Fig2}
\end{figure}

\section{Enantiodiscrimination of chiral molecules\label{detchirsec}}
In Sec.~\ref{corrfunsec}, we have calculated analytically the equal-time second-order correlation function of the cavity field relying on the overall phase. In this section, we will discriminate the left- and right-handed chiral molecules by detecting the correlation function of the cavity field. In the following numerical simulations, we consider that the molecular transitions $\vert2\rangle_{Q}\leftrightarrow\vert3\rangle_{Q}$ and $\vert1\rangle_{Q}\leftrightarrow\vert2\rangle_{Q}$ are, respectively, resonantly coupled with the classical light field and quantized cavity field, i.e., $\Delta_{32}=0$ and $\Delta_{31}=\Delta_{c}$. In addition, we employ the experimentally feasible parameters: $\kappa/2\pi=1$\,MHz~\cite{Kampschulte2018Cavity,Hoghooghi2019Broadband}, $\gamma_{21}/2\pi=\gamma_{31}/2\pi=\gamma_{32}/2\pi=0.1$\,MHz~\cite{Patterson2013Enantiomer,Patterson2013Sensitive}, $g/2\pi=0.1$\,MHz~\cite{Shalabney2015Coherent,Long2015Coherent,Vergauwe2016Quantum}, $\Omega_{32}/2\pi=(0.1\sim1)$\,MHz, $\Omega_{31}/2\pi=(0\sim0.03)$\,MHz, and $\xi_{p}/2\pi=0.01$\,MHz. Since these driving strengths ($\Omega_{32}$, $\Omega_{31}$, and $\xi_{p}$) are tunable on demand, all the conditions of the parameters can be fulfilled simultaneously in a cavity-molecule system for current experimental technology.

In Fig.~\ref{Fig2}, we show the equal-time second-order correlation function $g^{(2)}_{Q}(0)$ ($Q=L,R$) as a function of the detuning $\Delta_{c}$. Concretely, we consider the cases of $\phi=0$ in Fig.~\ref{Fig2}(a) and $\phi=\pi/2$ in Fig.~\ref{Fig2}(b). Here the solid curves correspond to the numerical results obtained by solving Eq.~(\ref{MEQ}), while the dashed curves correspond to the analytical results given in Eq.~(\ref{corrfun}). We find that there is a slight discrepancy between the analytical and numerical results, but the main physical results are the same. Such a discrepancy can be understood as follows: (i) We have mentioned in Sec.~\ref{corrfunsec} that the quantum jump terms are ignored in the analytical calculation compared with the numerical result. (ii) In the derivation of the analytical result, we have made the perturbation approximation. In the numerical calculation, the quantum master equation~(\ref{MEQ}) is solved with the numerical method. (iii) In the derivation of the analytical result, we choose the truncation dimension of the cavity field as $n_{c}=2$. However, in the numerical simulation, the appropriate truncation dimension of the cavity field needs to be chosen such that the numerical result of the correlation function is convergent. Hence, here we choose the truncation dimension of the cavity field as $n_{c}=8$. In the case of $\phi=0$, we can see from Fig.~\ref{Fig2}(a) that the correlation function $g^{(2)}_{L}(0)$ for the left-handed chiral molecule is larger than $1$ around $\Delta_{c}=0$ (the light orange area), which corresponds to the super-Poissonian distribution of the photons. However, the correlation function for the right-handed chiral molecule is $g^{(2)}_{R}(0)<1$ around $\Delta_{c}=0$ corresponding to the sub-Poissonian distribution. For the case of $\phi=\pi/2$, it can be seen in Fig.~\ref{Fig2}(b) that, in the red-detuning (blue-detuning) regime $\Delta_{c}>0$ ($\Delta_{c}<0$), a bunching peak of the correlation function $g^{(2)}_{L}(0)>1$ [$g^{(2)}_{R}(0)>1$] can be observed for the left-handed (right-handed) chiral molecule. In addition, we find that the correlation function for the left-handed chiral molecule is $g^{(2)}_{L}(0)>1$ [$g^{(2)}_{L}(0)<1$] and for the right-handed chiral molecule is $g^{(2)}_{R}(0)<1$ [$g^{(2)}_{R}(0)>1$] in the red (blue) area. This indicates that the left- and right-handed chiral molecules can be discriminated by detecting the equal-time second-order correlation function of the cavity field.
\begin{figure}
\center
\includegraphics[width=0.65 \textwidth]{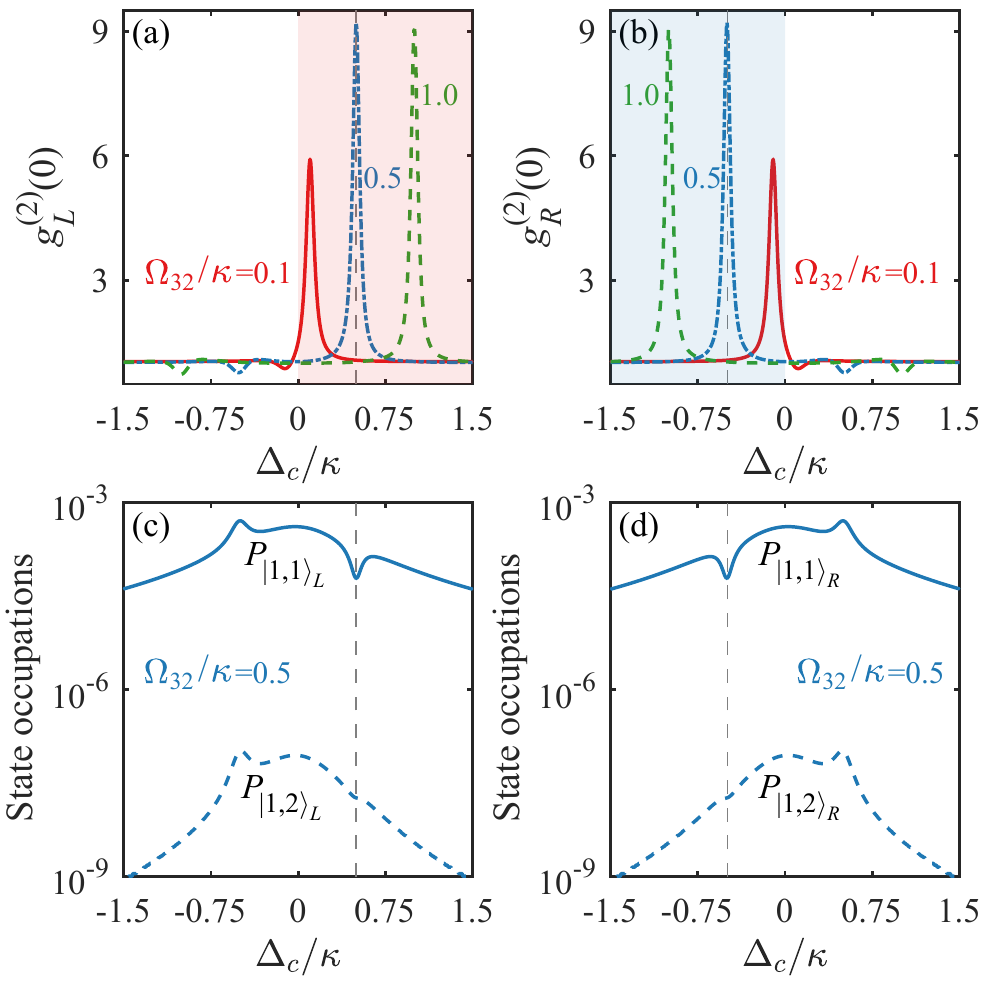}
\caption{The correlation functions (a) $g^{(2)}_{L}(0)$ and (b) $g^{(2)}_{R}(0)$ for the left- and right-handed chiral molecules as functions of the detuning $\Delta_{c}$ at different values of $\Omega_{32}$. The steady-state occupations (c) $P_{\vert 1,1\rangle_{L}}=\,_{L}\langle1,1\vert\rho^{L}_{\text{ss}}\vert 1,1\rangle_{L}$ and $P_{\vert 1,2\rangle_{L}}=\,_{L}\langle1,2\vert\rho^{L}_{\text{ss}}\vert 1,2\rangle_{L}$ and (d) $P_{\vert 1,1\rangle_{R}}=\,_{R}\langle1,1\vert\rho^{R}_{\text{ss}}\vert 1,1\rangle_{R}$ and $P_{\vert 1,2\rangle_{Q}}=\,_{R}\langle1,2\vert\rho^{R}_{\text{ss}}\vert 1,2\rangle_{R}$ for the left- and right-handed chiral molecules versus the detuning $\Delta_{c}$ at $\Omega_{32}/\kappa=0.5$. Other parameters are chosen as $\Delta_{32}=0$, $\Delta_{31}=\Delta_{c}$, $\phi=\pi/2$, $\kappa/2\pi=1$\,MHz, $\gamma_{21}/2\pi=\gamma_{31}/2\pi=\gamma_{32}/2\pi=0.1$\,MHz, $\gamma_{\phi,21}=\gamma_{\phi,31}=\gamma_{\phi,32}=0$, $g/2\pi=0.1$\,MHz, and $\xi_{p}/2\pi=\Omega_{31}/2\pi=0.01$\,MHz.}
\label{Fig3}
\end{figure}

To analyze the influence of the driving strength $\Omega_{32}$ of the classical light field acting on the molecular transition $\vert2\rangle_{Q}\leftrightarrow\vert3\rangle_{Q}$ on the correlation function, we display in Figs.~\ref{Fig3}(a) and~\ref{Fig3}(b) the correlation functions $g^{(2)}_{Q}(0)$ for the left- and right-handed chiral molecules versus the detuning $\Delta_{c}$ at various values of $\Omega_{32}$. All results in Fig.~\ref{Fig3} are obtained by solving Eq.~(\ref{MEQ}), and we consider only the case of $\phi=\pi/2$. It can be seen that the detuning locations of the bunching peaks in the correlation function for the left- and right-handed chiral molecules correspond to the red detuning $\Delta_{c}\approx\Omega_{32}$ and the blue detuning $\Delta_{c}\approx-\Omega_{32}$, respectively. In particular, we find that the curve of the correlation function for the left-handed chiral molecule is mutually symmetric with the right-handed chiral molecule at $\phi=\pi/2$. To further understand the reason of the bunching peak generation, we also analyze the steady-state occupations $P_{\vert 1,1\rangle_{Q}}=\,_{Q}\langle1,1\vert\rho^{Q}_{\text{ss}}\vert 1,1\rangle_{Q}$ and $P_{\vert 1,2\rangle_{Q}}=\,_{Q}\langle1,2\vert\rho^{Q}_{\text{ss}}\vert 1,2\rangle_{Q}$ for the left- and right-handed chiral molecules versus the detuning $\Delta_{c}$ at $\Omega_{32}/\kappa=0.5$, as shown in Figs.~\ref{Fig3}(c) and~\ref{Fig3}(d). We find that the location of the bunching peak in the correlation function $g^{(2)}_{Q}(0)$ corresponds to the location of the dip in $P_{\vert 1,1\rangle_{Q}}$. For the cavity-molecule system, there are two different transition paths from the state $\vert1,0\rangle_{Q}$ to $\vert1,1\rangle_{Q}$. Here the direct transition path is $\vert1,0\rangle_{Q}\overset{\xi_{p}}{\longrightarrow}\vert1,1\rangle_{Q}$, and the indirect transition path is $\vert1,0\rangle_{Q}\overset{\Omega_{31}}{\longrightarrow}\vert3,0\rangle_{Q}\overset{\Omega_{32}e^{i\phi_{Q}}}{\longrightarrow}\vert2,0\rangle_{Q}
\overset{g}{\longrightarrow}\vert1,1\rangle_{Q}$. When a perfect destructive quantum interference happens between these two transition paths, the value of the steady-state occupation $P_{\vert 1,1\rangle_{Q}}$ is zero. However, the value of $P_{\vert 1,1\rangle_{Q}}$ decreases for an imperfect destructive quantum interference, corresponding to the location of the dip in $P_{\vert 1,1\rangle_{Q}}$. This indicates that the generation of the bunching peak is based on the quantum interference effect between the two different transition paths~\cite{Zou2020Enhancement}.

\begin{figure}
\center
\includegraphics[width=0.65 \textwidth]{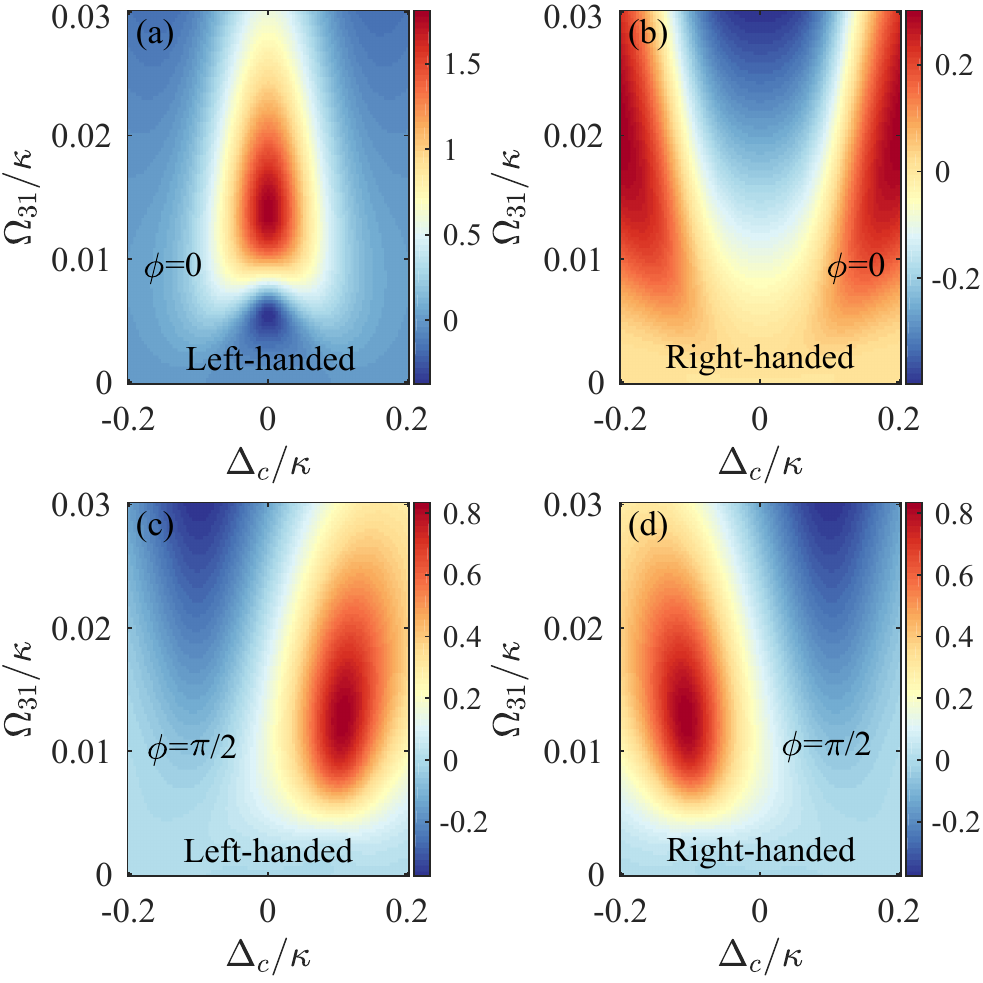}
\caption{The correlation function $\text{log}_{10}g^{(2)}_{L}(0)$ for the left-handed chiral molecule [$\text{log}_{10}g^{(2)}_{R}(0)$ for the right-handed chiral molecule] as a function of $\Delta_{c}$ and $\Omega_{31}$ for (a) $\phi=0$ and (c) $\phi=\pi/2$ [(b) $\phi=0$ and (d) $\phi=\pi/2$]. Other parameters are chosen as $\Delta_{32}=0$, $\Delta_{31}=\Delta_{c}$, $\kappa/2\pi=1$\,MHz, $\gamma_{21}/2\pi=\gamma_{31}/2\pi=\gamma_{32}/2\pi=0.1$\,MHz, $\gamma_{\phi,21}=\gamma_{\phi,31}=\gamma_{\phi,32}=0$, $g/2\pi=0.1$\,MHz, $\Omega_{32}/2\pi=0.1$\,MHz, and $\xi_{p}/2\pi=0.01$\,MHz.}
\label{Fig4}
\end{figure}

We also analyze how the correlation function depends on the driving strength $\Omega_{31}$ of the classical light field acting on the molecular transition $\vert1\rangle_{Q}\leftrightarrow\vert3\rangle_{Q}$. The correlation functions $\text{log}_{10}g^{(2)}_{Q}(0)$ ($Q=L,R$) are plotted versus the detuning $\Delta_{c}$ and the driving field $\Omega_{31}$ at $\phi=0$ [in Figs.~\ref{Fig4}(a) and~\ref{Fig4}(b)] and $\phi=\pi/2$ [in Figs.~\ref{Fig4}(c) and~\ref{Fig4}(d)]. In the case of $\phi=0$, we see in Figs.~\ref{Fig4}(a) and~\ref{Fig4}(b) that the correlation function $g^{(2)}_{L}(0)$ [$g^{(2)}_{R}(0)$] is larger (less) than $1$ around $\Delta_{c}=0$ when the driving strength $\Omega_{31}/\kappa\in[0.01,0.03]$, which means that the left- and right-handed chiral molecules can be discriminated under the condition of $0.01\leq\Omega_{31}/\kappa\leq0.03$. For $\phi=\pi/2$, it can be observed from Figs.~\ref{Fig4}(c) and~\ref{Fig4}(d) that the left- and right-handed chiral molecules can be distinguished at the range of $0.005\leq\Omega_{31}/\kappa\leq0.03$. From the above analysis, it can be found that, under the appropriate parameter conditions, discrimination between the left- and right-handed chiral molecules can be accomplished by detecting the equal-time second-order correlation function of the cavity field.

\begin{figure}
\center
\includegraphics[width=0.65 \textwidth]{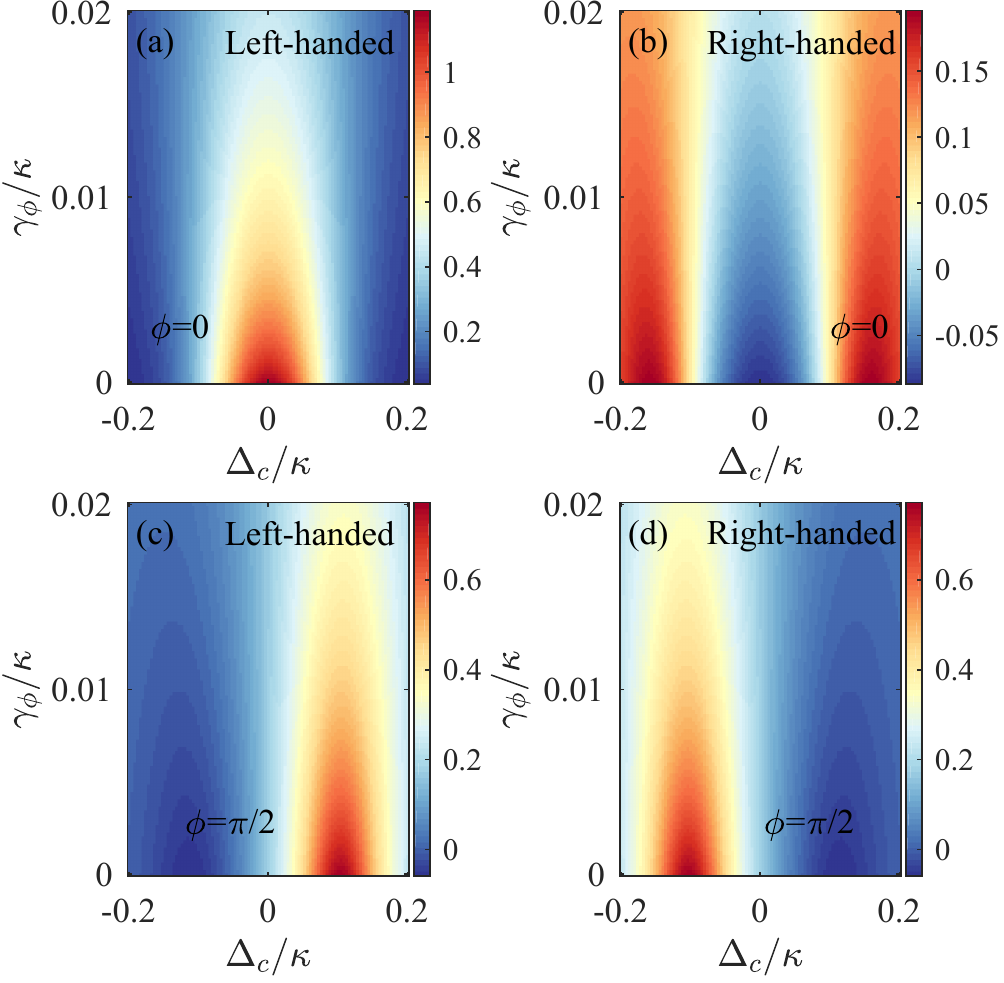}
\caption{The correlation function $\text{log}_{10}g^{(2)}_{L}(0)$ for the left-handed chiral molecule [$\text{log}_{10}g^{(2)}_{R}(0)$ for the right-handed chiral molecule] as a function of $\Delta_{c}$ and $\gamma_{\phi}$ for (a) $\phi=0$ and (c) $\phi=\pi/2$ [(b) $\phi=0$ and (d) $\phi=\pi/2$]. Other parameters are chosen as $\Delta_{32}=0$, $\Delta_{31}=\Delta_{c}$, $\kappa/2\pi=1$\,MHz, $\gamma_{21}/2\pi=\gamma_{31}/2\pi=\gamma_{32}/2\pi=0.1$\,MHz, $\gamma_{\phi,21}=\gamma_{\phi,31}=\gamma_{\phi,32}=\gamma_{\phi}$, $g/2\pi=0.1$\,MHz, $\Omega_{32}/2\pi=0.1$\,MHz, and $\xi_{p}/2\pi=\Omega_{31}/2\pi=0.01$\,MHz.}
\label{Fig5}
\end{figure}

In the above discussions, we neglect the effect of the pure dephasing rates in the system. Below, we will discuss the influence of the pure dephasing rates on the correlation function. Here we consider the case of $\gamma_{\phi,21}=\gamma_{\phi,31}=\gamma_{\phi,32}=\gamma_{\phi}$. The correlation functions $\text{log}_{10}g^{(2)}_{Q}(0)$ ($Q=L,R$) are plotted versus the detuning $\Delta_{c}$ and the pure dephasing rate $\gamma_{\phi}$ at $\phi=0$ [in Figs.~\ref{Fig5}(a) and~\ref{Fig5}(b)] and $\phi=\pi/2$ [in Figs.~\ref{Fig5}(c) and~\ref{Fig5}(d)].  In the case of $\phi=0$, it can be seen from Figs.~\ref{Fig5}(a) and~\ref{Fig5}(b) that the correlation function $g^{(2)}_{L}(0)$ [$g^{(2)}_{R}(0)$] is larger (less) than 1 around $\Delta_{c}=0$ when the pure dephasing rate $\gamma_{\phi}\in[0,0.01]$. In the case of $\phi=\pi/2$, one can see from Figs.~\ref{Fig5}(c) and~\ref{Fig5}(d) that the bunching peak of the correlation function for the left-handed (right-handed) chiral molecule can be observed at $\Delta_{c}\approx\Omega_{32}$ ($\Delta_{c}\approx-\Omega_{32}$) when $\gamma_{\phi}/\kappa\in[0,0.02]$. This indicates that the left- and right-handed chiral molecules can also be discriminated under the appropriate parameter conditions when considering the pure dephasing of the system. In addition, it can be found from Figs.~\ref{Fig4} and~\ref{Fig5} that our scheme of enantiodiscrimination is robust against the fluctuation of the parameters under the appropriate parameter conditions (though the value of the equal-time second-order correlation function may be not robust against the fluctuation of the parameters).

Finally, we present a discussion on the implementation of the scheme. Concretely, we consider 1,2-propanediol molecules~\cite{Ayrea1994The,Ayrea2017A,lovas2009Microwave} as an example to realize the single-loop three-level models. We choose three working states of the cyclic three-level model as $\vert1\rangle=\vert g\rangle\vert0_{000}\rangle$, $\vert2\rangle=\vert e\rangle\vert1_{110}\rangle$, and $\vert3\rangle=\vert e\rangle(\vert1_{101}\rangle+\vert1_{10-1}\rangle)/\sqrt{2}$, where $\vert g\rangle$ and $\vert e\rangle$ are, respectively, the vibrational ground state and first-excited state with transition frequency $\omega_{\text{vib}}=2\pi\times100.950$\,THz~\cite{Ayrea1994The}. The rotational state is marked as $\vert J_{K_{a}K_{c}M}\rangle$~\cite{Leibscher2019Principles,Zare1998Angular} with the angular moment quantum number $J$, the magnetic quantum number $M$, and $K_{a}$ ($K_{c}$) runs form $J$ (0) to 0 ($J$) in unit step with decreasing energy. According to the rotational
constants for 1,2-propanediol molecules $A=2\pi\times8524.405$\,MHz, $B=2\pi\times3635.492$\,MHz, and $C=2\pi\times2788.699$\,MHz~\cite{Ayrea2017A}, we can obtain the transition frequencies between the three working states as $\omega_{21}\equiv\omega_{2}-\omega_{1}=2\pi\times100.961$\,THz, $\omega_{31}\equiv\omega_{3}-\omega_{1}=2\pi\times100.962$\,THz, and $\omega_{32}\equiv\omega_{3}-\omega_{2}=2\pi\times0.847$\,GHz~\cite{Zare1998Angular}. Hence, the single-loop three-level models can be realized by choosing three linearly polarized electromagnetic fields. Here the state $\vert1\rangle$ is coupled to the state $\vert2\rangle$ by the $z$-polarized quantized light field, and the state $\vert1\rangle$ ($\vert2\rangle$) is coupled to the state $\vert3\rangle$ by the $y$-polarized ($x$-polarized) classical light field based on electric-dipole interaction. In particular, the frequencies of the three electromagnetic fields can be all microwave fields~\cite{Patterson2013Enantiomer,Patterson2013Sensitive,
Patterson2014New,Shubert2014Identifying,Shubert2015Rotational,Shubert2016Chiral,Lobsiger2015Molecular}, or one microwave and two infrared fields~\cite{Chen2021Enantio,Chen2021arXivEnantio,Zhang2020Evading}. For the case of the microwave (infrared) cavity with quantized field, it can be a transmission line resonator (Fabry-P\'{e}rot cavity).

\section{Conclusion\label{conclusion}}
In conclusion, we have proposed a feasible scheme to discriminate molecular chirality by detecting the equal-time second-order correlation function of the cavity field in a driven cavity-molecule system. By analytically and numerically calculating the correlation function, we find that this quantum correlation function for the left-handed (right-handed) chiral molecule is $g^{(2)}_{L}(0)>1$ [$g^{(2)}_{R}(0)<1$] around $\Delta_{c}=0$ for the overall phase $\phi=0$. In the case of $\phi=\pi/2$, we find that the bunching peak of the second-order correlation function for the left-handed (right-handed) chiral molecule can be observed in the red-detuning (blue-detuning) regime. The obtained results indicate that the left- and right-handed chiral molecules can be discriminated by detecting the equal-time second-order correlation function. Our work provides a promising method to discriminate molecular chirality and has the following features: (i) For the case of single molecule, when the single-molecule coupling is weak (e.g. the coupling strength between the single molecule and the quantum light field is less than the decay rate of cavity field), the left- and right-handed chiral molecules can also be discriminated by measuring the correlation function of the cavity field. For the single-molecule case with weak coupling, the previous scheme (e.g. Refs.~\cite{Chen2021Enantio,Chen2021arXivEnantio}) would not work well. (ii) Our scheme of enantiodiscrimination is robust against the fluctuation of the parameters under the appropriate parameter conditions.

\begin{backmatter}

\bmsection{Funding}
This work is supported in part by the National Natural Science Foundation of China (Grants No.~12074030, No.~12147109, No.~11774024, and No.~U1930402), and the China Postdoctoral Science Foundation (Grant No.~2021M690323 and No.~2021M700360).

\bmsection{Disclosures}
The authors declare no conflicts of interest.

\bmsection{Data Availability}
Data underlying the results presented in this paper are not publicly available at
this time but may be obtained from the authors upon reasonable request.

\end{backmatter}


\end{document}